\documentclass[journal]{IEEEtran}
\ifCLASSINFOpdf

\else

\fi

\usepackage{amsmath}
\usepackage{makeidx}  
\usepackage{algorithm}
\usepackage{algorithmic}
\usepackage{graphicx}
\usepackage{subfigure}
\usepackage{epstopdf}
\usepackage{bm}
\usepackage{cite}
\usepackage{stfloats}

\usepackage{amssymb}
\usepackage{graphicx}

\usepackage{url}
\newtheorem{proposition}{Proposition}
\usepackage{tabularx,booktabs}
\newcolumntype{C}{>{\centering\arraybackslash}X} 
\usepackage{lipsum}

\usepackage{makecell} 

\usepackage{graphicx}
\usepackage{color}
\newtheorem{thm}{Theorem}

\newtheorem{remark}{Remark}

\begin{document}

\title{Uplink Achievable Rate of  Intelligent Reflecting Surface-Aided  Millimeter-Wave Communications with Low-Resolution ADC and Phase Noise}

\author{Kangda Zhi, Cunhua Pan, Hong Ren and Kezhi Wang

	\thanks{(Corresponding author: Cunhua Pan).
		
		K. Zhi, C. Pan are with the School of Electronic Engineering and Computer Science at Queen Mary University of London, London E1 4NS, U.K. (e-mail: k.zhi, c.pan@qmul.ac.uk).
		
		H. Ren is with the National Mobile Communications Research Laboratory, Southeast University, Nanjing 210096, China. (hren@seu.edu.cn).
		
		K. Wang is with Department of Computer and Information Sciences, Northumbria University, UK. (e-mail: kezhi.wang@northumbria.ac.uk).}}

\maketitle

\begin{abstract}
In this paper, we derive the uplink achievable rate expression of  intelligent reflecting surface (IRS)-aided millimeter-wave (mmWave) systems, taking into account the phase noise at IRS and the quantization error at base stations (BSs). We show that the performance is limited only by the resolution of analog-digital converters (ADCs) at BSs when the number of IRS reflectors grows without bound. On the other hand, if BSs have ideal ADCs, the performance loss  caused by IRS phase noise is constant. Finally, our results validate the feasibility of using low-precision hardware at the IRS  when BSs are equipped with low-resolution ADCs.
\end{abstract}
%
%
%

\section{Introduction}
Millimeter-wave (mmWave) technology will play an important role in future networks, because of its capability of achieving high system capacity, increased security and reduced interference{\cite{JSACmmwave2017}. However, blockage issue needs to be tackled before the commercial application of this technology. In specific, mmWave frequencies are susceptible to the blockage, which means  mmWave communications are difficult to be applied in urban areas with dense buildings and vehicles. 
	
To cope with this issue, intelligent reflecting surface (IRS), also referred to as reconfigurable intelligent surface (RIS), has been proposed as an attractive technology to create virtual line-of-sight (LoS) links for mmWave communication systems{\cite{di2020smart}}. IRS is composed of a large number of reconfigurable passive elements, which is able to achieve passive beamforming  by adjusting the phase of  impinging signal. Moreover, since IRS can avoid the use of radio-frequency (RF) chains, it can  effectively reduce the hardware cost and energy consumption compared with conventional active relay nodes {\cite{wu2020intelligent,9090356,8811733,9110849}}.

{Since realizing fully digital architecture in mmWave systems requires prohibitive hardware cost and power consumption\cite{JSACmmwave2017},  IRS-aided mmWave systems with the hybrid analog-digital architecture have been investigated in some prior works\cite{zhou2020stochastic,TVTwang,pradhan2020hybrid}. Specifically, the authors in {\cite{zhou2020stochastic}} demonstrated the promising reliability and connectivity of IRS-aided mmWave systems with random blockages.} The joint  passive and active beamforming optimization for  single-user mmWave networks was studied in {\cite{TVTwang}}, in which closed-form solution was obtained for the single-IRS case and low-complexity iterative algorithm was proposed for the multiple-IRS case.  The authors in {\cite{pradhan2020hybrid}}  jointly optimized hybrid precoding at BSs and phase shifting at the IRS to minimize the mean-squared error (MSE) for multiuser mmWave systems.

{In addition to the hybrid analog-digital precoding, another promising solution for reducing RF complexity in mmWave systems is low-resolution digital-to-analog converters (DACs) and analog-to-digital converters (ADCs)\cite{JSACmmwave2017}. Moreover, compared with the DACs at the transmitter, much more power consumption is needed for the ADCs at the receiver side \cite{6979963}. Therefore, it is meaningful to study the performance of IRS-aided mmWave systems with low-resolution ADCs. However, to the best of our knowledge, none of the work has considered this scenario. On the other hand, IRS phase noise, caused by its discrete phase shifts, has been recognized as an essential factor in practical IRS-aided systems}{\cite{badiu2019communication,zhou2020spectral,han2019large,li2020ergodic,xing2020achievable}}. Considering the phase estimation and  quantization errors, the authors in {\cite{badiu2019communication}} derived the distribution of signal-to-noise ratio (SNR) and the average error probability in single-antenna single-user systems. In downlink multiple-input single-output (MISO) systems,  the effects of phase noise at IRS and RF impairments at BSs were  studied in {\cite{zhou2020spectral}}. Under the assumption of  Rician fading channel model, the minimum number of   quantization phase bits to ensure a performance  loss threshold was derived in {\cite{han2019large}} by using the upper bound of the ergodic spectral efficiency.  The impact of phase error on the rate loss was quantified in {\cite{li2020ergodic}}  for an IRS-aided single-antenna system with the existence of direct link. Considering the phase errors at the IRS and the distortion noise at  the BS, the authors in {\cite{xing2020achievable}}  investigated the rate performance of  IRS-aided networks and then theoretically compared it with the relay-aided networks.

{Motivated by the above discussions, we focus on the uplink IRS-aided mmWave systems with the existence of both low-resolution ADCs and IRS phase noise, where the IRS is used to create an additional virtual  LoS link for the blocked user. We aim to theoretically evaluate the capacity performance of this system and characterize the joint impacts of low-resolution ADCs and IRS phase noise.} Specifically, we first derive the closed-form expression of the uplink achievable rate. Then, based on the derived expression, we analyze the performance loss caused by the  phase noise at the IRS and  quantization error at the BS. Our asymptotic  results reveal that when the number of IRS phase shifts approaches infinity, the achievable rate is limited by the precision of the ADCs at BSs. In addition, our results demonstrate the feasibility of utilizing simple IRSs with a few phase shift levels when BSs are equipped with low-resolution ADCs.

\section{System Model}
\begin{figure}[t]
\setlength{\abovecaptionskip}{0pt}
\setlength{\belowcaptionskip}{-20pt}
\centering
\includegraphics[width= 0.25\textwidth]{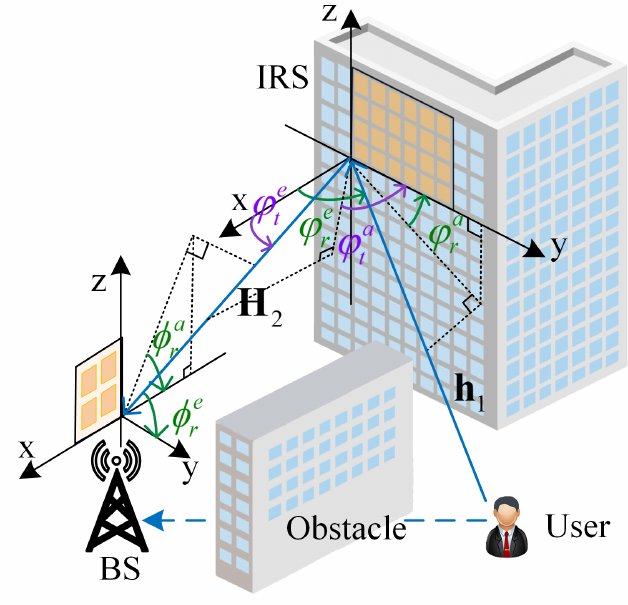}
\DeclareGraphicsExtensions.
\caption{An IRS-assisted uplink mmWave communication system.}
\label{figure0}
\vspace{-10pt}
\end{figure}
We consider the uplink mmWave communication of an IRS-aided system with an $M$-antenna BS, an IRS consisting of $N$ reflecting elements and a single-antenna user, as shown in Fig. \ref{figure0}. Considering the blockage sensitivity in the mmWave frequencies, we assume that the direct link is blocked by the building. To improve the quality of service,  an IRS is deployed in a proper place such that there exists an alternative user-IRS-BS link.

We assume that  both the IRS and the BS are equipped with the uniform square planar array (USPA), and the BS can control the phase shift of each IRS's  reflecting  element. The phase shift matrix of the IRS can then be given by
\begin{align}\label{IRS3}
{\bf \Theta } = {\rm{diag}}\left\{ {{\zeta _1}{e^{j{\theta _1}}},{\zeta _2}{e^{j{\theta _2}}},...,{\zeta _N}{e^{j{\theta _N}}}} \right\},
\end{align}
where ${{\zeta _n}} \in \left[ {0,1} \right]$ and $\theta _n \in \left[ {0,2\pi} \right),\forall n$ denote the {reflection amplitude} and the phase shift induced by the $n$th reflecting element  of the IRS, respectively. As the IRS is designed  to enhance the reflected signal's quality, we set ${{\zeta _n}} =1,\forall n $.{\footnote{{ To facilitate our analysis, we ignore the non-linear coupling between the reflection amplitude and phase shifts which has been discussed in \cite{practical2020,wu2020intelligent}. This practical model will be considered in our future work.}}}Because of the hardware limitation,  IRS  can only  choose its phase shifts from a finite number of discrete values. Therefore, we assume the number of the quantization bits for the phase shifts at the IRS is $B$, and the set of discrete phase shifts available at the IRS can be denoted by $\left\{   0,{\frac{2\pi}{2^B}},...,(2^B-1){\frac{2\pi}{2^B}}    \right\} $.
Then, the phase noise at IRS reflector $n$ can be written as ${\hat \theta _n} \sim {\mathcal U} \left[ \frac{-\pi}{2^B},\frac{\pi}{2^B} \right],\forall n$,
where ${\mathcal U}$ denotes the uniform distribution. Therefore, the phase shift matrix of the IRS with phase noise can be expressed as
\begin{align}\label{IRS4}
\tilde {\bf \Theta}  = {\rm{diag}}\left\{ {{\zeta _1}{e^{j{{\tilde \theta }_1}}},{\zeta _2}{e^{j{{\tilde \theta }_2}}},...,{\zeta _N}{e^{j{{\tilde \theta }_N}}}} \right\},
\end{align}
where ${\tilde \theta _n} = {\theta _n} + {\hat \theta _n}$ denotes the actual phase shift of the IRS element $n$ with $\theta_n$ being the designed phase shift to be detailed later. 

{By properly deploying the IRS, the virtual LoS channel created by IRS in mmWave systems can be strongly sparse. Thus, we adopt the widely used geometry channel model \cite{zhou2020stochastic,TVTwang,pradhan2020hybrid} with only one dominant propagation path}. The channel from the user to the IRS and that from the IRS to the BS can be respectively expressed as
\begin{align}\label{IRS1}
{\bf h_1} = \alpha {{\bf a}_N}\left( {\varphi _r^a,\varphi _r^e} \right), {\bf H_2} = \beta {{\bf a}_M}\left( {\phi _r^a,\phi _r^e} \right){\bf a}_N^H\left( {\varphi _t^a,\varphi _t^e} \right),
\end{align}
where $\alpha$ and $\beta$ are the link strengths. {As shown in Fig. \ref{figure0},} $\varphi _r^a,\varphi _r^e$ ($\phi _r^a,\phi _r^e$) are the azimuth and elevation angles of arrival (AOA) at the IRS (BS), $\varphi _t^a,\varphi _t^e$ are the azimuth and elevation angles of departure (AoD) at the IRS. ${{\bf a}_X}\left( {\vartheta _{}^a,\vartheta _{}^e} \right)$ is the array response vector of the USPA with size $\sqrt X  \times \sqrt X $, which could be written as
{\begin{align}\label{IRS2}
\begin{array}{l}
{{\bf a}_X}\left( {\vartheta _{}^a,\vartheta _{}^e} \right) \!= [1,...,{e^{j2\pi \frac{d}{\lambda }\left( {x\sin \vartheta _{}^e\sin \vartheta _{}^a + y\sin \vartheta _{}^e\cos \vartheta _{}^a} \right)}},\\
...,{e^{j2\pi \frac{d}{\lambda }\left( {\left( {\sqrt X  - 1} \right)\sin \vartheta _{}^e\sin \vartheta _{}^a + \left( {\sqrt X  - 1} \right)\sin \vartheta _{}^e\cos \vartheta _{}^a} \right)}}{]^T},
\end{array}
\end{align}
}where $\lambda$ and $d$ are the carrier wavelength and element spacing, respectively. ${\rm{0}} \le x,y \le \sqrt X  - 1$ are the element indices of the USPA. In this paper, we assume that angles in (\ref{IRS1}) can be perfectly estimated based on some channel estimation methods\cite{wang2020joint,deepak2020channel} and the case with imperfect angular information as in \cite{2020uncertainCSi} will be left for our future work.

%

To reduce the  power consumption and hardware cost in mmWave band,  low-resolution ADCs are adopted at the BS. For analytical tractability, we adopt the additive quantization noise model (AQNM) to characterize the impacts of the low-resolution ADCs at the BSs. Thus, the quantizer outputs of the received signal with phase noise can be written as {\cite{fan2015uplink}}
\begin{align}\label{IRS6}
{{\bf y}_q} = \gamma {\bf y} + {{\bf n}_q} = \gamma \left({\sqrt P \tilde {\bf g}{s} + {\bf n}} \right)+ {{\bf n}_q}= \gamma \sqrt P \tilde {\bf g}{s} + \gamma {\bf n} + {{\bf n}_q},
\end{align}
where $\tilde {\bf g} = {\bf H_2}\tilde {\bf \Theta} {\bf h_1}$ represents the $M \times 1$ cascaded user-IRS-BS channel with phase noise, ${s}$ denotes the transmit symbol, ${\bf n} $ is a normalized additive white Gaussian noise vector with zero mean and unit variance. Therefore, $P$ represents the normalized transmit SNR. Besides, $\gamma = 1-\rho $, where $\rho$ is the inverse of the signal-to-quantization-noise ratio and ${\bf n}_q$ denotes the additive Gaussian quantization noise vector which is uncorrelated with $\bf y$. The covariance matrix of ${\bf n}_q$ is given by
\begin{align}
{{\bf R}_{{{\bf n}_q}{{\bf n}_q}}} = \gamma \left( {1 - \gamma } \right) {\rm diag} \left( {P\tilde {\bf g}{{\tilde {\bf g} }^H} + {\bf I}} \right).
\end{align}

Considering the non-uniform scalar minimum mean-square-error (MMSE) quantizer of the Gaussian random variables, the values of $\rho$ corresponding to the quantization bits $b$ can be found from {Table \ref{tab1}}{\cite{fan2015uplink}}.
\vspace{-10pt}
\begin{table}[H]
\centering
\caption{$\rho$ versus Quantization Bits $b$}
\vspace{-5pt}
\begin{tabular}{c|c|c|c|c|c|c}
               \hline
               $b$ &1  &2 &3 &4 & 5&$ \ge 6$\\
               \hline
                $\rho$ &0.3634&0.1175&0.03454&0.009497&0.002499&$\frac{{\sqrt 3 \pi }}{2}{2^{ - 2b}}$\\
               \hline
 \end{tabular}\label{tab1}
\end{table}

\section{Analysis of Achievable Uplink Rate}
In this section, we analyze the uplink achievable rate and investigate the effect of low-resolution ADCs at the BS and phase noise at the IRS. In addition, we  characterize the performance gap between our results and the ideal case without quantization error or phase noise.

Consider the maximal-ratio-combining (MRC) receiver. Since the phase noise is unknown, the receive beamforming is designed based on the available channel state information (CSI) and on the optimized phase shift of the IRS. Thus, the quantized signal received by the BS can be processed as
\begin{align}
{ r} = {{\bf g}^H}{{\bf y}_q} = \gamma \sqrt P {{\bf g}^H}\tilde {\bf g}{s}+ \gamma {{\bf g}^H}{\bf n} + {{\bf g}^H}{{\bf n}_q},
\end{align}
where $ {\bf g} = {\bf H_2} {\bf \Theta} {\bf h_1}$.

{Then the uplink achievable rate can be written as
\begin{align}\label{rate}
R={\rm log}_2 \!\left( 1+\frac{{{\gamma ^2}P{{\left| {{{\bf g}^H}\tilde {\bf g}} \right|}^2}}}{{{\gamma ^2}{{\left\| {\bf g} \right\|}^2} +     \gamma \left( {1 - \gamma } \right) {\bf g}^H{\rm diag} \left( {P\tilde {\bf g}{{\tilde {\bf g} }^H} + {\bf I}} \right){\bf g}}}\right).
\end{align}}}

\subsection{ Phase Shift Matrix Design}
{Since the phase noise at the IRS is unknown, we maximize the SNR by designing the phase shifts without considering the existence of phase noise. Although this design is sub-optimal, it has lower complexity in practical systems and can lead to a tractable analytical expression. The optimization problem can be formulated as follows
	\begin{align}
	\begin{array}{l}
	\max\limits_{\bf \Theta}  \quad {\rm SNR}^ {ideal }=\frac{\gamma^{2} P\left|\mathbf{g}^{H} \mathbf{g}\right|^{2}}{\gamma^{2}\|\mathbf{g}\|^{2}+\gamma(1-\gamma) \mathbf{g}^{H} \operatorname{diag}\left(P \mathbf{g g}^{H}+\mathbf{I}\right) \mathbf{g}}, \\
	\;{\rm s.t. } \;\;\quad { \theta }_{n} \in[0,2 \pi), \forall n.
	\end{array}
	\end{align}
	
	Recalling (\ref{IRS1}), ideal cascaded channel $\bf g$ can be rewritten as
	\begin{align}
	\begin{array}{l}
	\mathbf{g}=\mathbf{H}_{2} \mathbf{\Theta} \mathbf{h}_{1}=\alpha \beta \mathbf{a}_{M}\left(\phi_{r}^{a}, \phi_{r}^{e}\right) f({\bf \Theta}), \\
	\left|{\rm g}_{i}\right|=\left|{\alpha \beta}\right| \left|f({\mathbf{\Theta}})\right|,
	\end{array}
	\end{align}
	where $ f(\mathbf{\Theta}) \triangleq \mathbf{a}_{N}^{H}\left(\varphi_{t}^{a}, \varphi_{t}^{e}\right) \mathbf{\Theta} \mathbf{a}_{N}\left(\varphi_{r}^{a}, \varphi_{r}^{e}\right)$ and $\mathrm{g}_i$ is the $i$-th entry of vector $\mathbf{g}$. Then, we can respectively calculate the three terms in $\mathrm{SNR}^{ideal}$ as follows
	\begin{align}\label{terms_ideal3}
	\begin{array}{l}
	\|\mathbf{g}\|^{2}=|\alpha \beta|^{2}\left\|\mathbf{a}_{M}\left(\phi_{r}^{a}, \phi_{r}^{e}\right)\right\|^{2}|f(\mathbf{\Theta})|^{2}=|\alpha \beta|^{2}M|f(\mathbf{\Theta})|^{2}, \\
	\left|\mathbf{g}^{H} \mathbf{g}\right|^{2}=\left(\|\mathbf{g}\|^{2}\right)^{2}=|\alpha \beta|^{4} M^{2}|f(\mathbf{\Theta})|^{4}, \\
	\mathbf{g}^{H} \operatorname{diag}\left(P{\mathbf{g}} \mathbf{g}^{H}+\mathbf{I}\right) \mathbf{g}=\sum_{i=1}^{M} {\mathrm{g}^{*}_i}\left(P {\mathrm{g}_i} {\mathrm{g}^{*}_i}+1\right) {\mathrm{g}_i} \\
	=\!\! \sum_{i=1}^{M}\!\! \left( \! P\left|{\mathrm{g}_i}\right|^{4} \!\!+\! \left|{\mathrm{g}_i}\right|^{2}\! \right) \!\!=\! M\! \left( \! P|\alpha \beta|^{4} |f(\mathbf{\Theta})|^{4} \!+\! |\alpha \beta|^{2}|f(\mathbf{\Theta})|^{2} \right).
	\end{array}
	\end{align}
	
	By substituting (\ref{terms_ideal3}) into $\mathrm{SNR}^{ideal}$, we can see that the maximal $\mathrm{SNR}^{ideal}$ is achieved under maximal  $|f(\mathbf{\Theta})|$. Thus, we have
	\begin{align}\label{optimal_PS}
	\begin{array}{l}
	{{\bf\Theta} ^{opt}} \!=\! \mathop {\arg \max }\limits_{\bf\Theta}  |f(\mathbf{\Theta})| \!=\!\mathop {\arg \max} \limits_{\bf\Theta}  {\left| {\sum_{n=1}^{N}  {{e^{j2\pi \frac{d}{\lambda }\left( {xp + yq} \right) + j{\theta _n}}}} } \!\right|},
	\end{array}
	\end{align}
	where $x = \lfloor {\left( {n - 1} \right)/\sqrt N} \rfloor $ and $y = \left( {n - 1} \right)\bmod \sqrt N $. $\left\lfloor n \right\rfloor $ means rounding $n$ toward the negative infinity and $\bmod$ means taking the remainder after division. $p = \sin \varphi _r^e\sin \varphi _r^a -  \sin \varphi _t^e\sin\varphi _t^a$ and $q = \sin \varphi _r^e\cos \varphi _r^a - \sin \varphi _t^e\cos \varphi _t^a$. Based on the triangle inequality,  the optimal phase shifts are given by
	\begin{align}\label{optimal_ps}
	\theta _n^{opt} =  - 2\pi \frac{d}{\lambda }\left( {x p + y q} \right), \forall n.
	\end{align}
}

\subsection{Uplink Rate Analysis}
Based on the optimal phase shift design in (\ref{optimal_ps}), we next derive the theoretical expression of the uplink achievable rate. The result is presented in the following theorem.
\begin{thm}\label{thm1}
Considering the impacts of the quantization error at the BSs and phase noise at the IRS, the uplink achievable rate for the IRS-assisted mmWave system with large $N$ is
\begin{align}\label{rateExp}
R \mathop  \to \limits^{a.s.} {\log _2}\left( {1 + \frac{{\gamma P{{\left| {\alpha \beta } \right|}^2}{N^2}M{{\rm sinc}^2}\left( \frac{\pi}{2^B} \right)}}{{\gamma  + \left( {1 - \gamma } \right)\left( {1 + P{{\left| {\alpha \beta } \right|}^2}{N^{\rm{2}}}{{{\mathop{\rm sinc}\nolimits} }^{\rm{2}}}\left( \frac{\pi}{2^B} \right)} \right)}}} \right).
\end{align}
\end{thm}

\itshape {Proof:}  \upshape See Appendix A. \hfill $\Box$

Theorem \ref{thm1} characterize the impacts of the number of IRS reflecting elements $N$, the number of antennas $M$ and transmit power $P$ on the uplink achievable rate when considering the quantization error at the BS and the phase noise at the IRS. The result in (\ref{rateExp}) shows that the uplink achievable rate scaling law is ${\mathcal O}(\log_2(M))$, which is not affected by the phase noise at the IRS. Besides, the following asymptotic results are presented to help us gain a better understanding of  Theorem \ref{thm1}.

\begin{remark}\label{remark11}
For the ideal ADC without quantization error, i.e., $\gamma  \to 1$, the uplink achievable rate  reduces to
\begin{align}\label{remark1}
R\mathop  \to \limits^{a.s.} {\log_2}\left( {1 + P{{\left| {\alpha \beta } \right|}^2}{N^2}M{\rm sinc^2}\left( \frac{\pi}{2^B} \right)} \right),
\end{align}
which shows that the achievable rate can increase with $N$ infinitely. Meanwhile, with a large $N$, (\ref{remark1}) can be approximated as
\begin{align}\label{remark1_1}
R \to {\log_2}\left( {P{{\left| {\alpha \beta } \right|}^2}{N^2}M} \right) + {\log_2}\left( {{{\rm sinc}^2}\left( \frac{\pi}{2^B}\right)} \right).
\end{align}
This indicates that the phase noise at the IRS will cause a constant {rate} loss ${\log_2}\left( {{{\rm sinc}^2}\left( \frac{\pi}{2^B} \right)} \right)$ when the number of  reflecting elements of the IRS becomes large.
\end{remark}

\begin{remark}\label{remark22}
With the quantization error and phase noise, when $P\!\! \to\! \!\infty $ or $N \!\!\to \!\!\infty $, the uplink achievable rate reduces to
\begin{align}\label{remark2}
R \to {\log _2}\left( {1 + \frac{{\gamma M}}{{\left( {1 - \gamma } \right)}}} \right),
\end{align}
which is independent of the phase noise at the IRS. Therefore, it is feasible to use simple hardware with high phase noise on the IRS when deploying a large number of reflecting elements or using high transmission power. In addition, (\ref{remark2}) shows that the achievable rate is mainly limited by the resolution of ADCs and cannot increase infinitely when $P$ or $N$ becomes infinite.
\end{remark}

Next, we give the following proposition to provide the design guidance for an uplink IRS-aided mmWave system with an acceptable performance degradation.
\begin{proposition}\label{proposition1}
	To design an uplink IRS-assisted mmWave system with rate degradation $\delta$, the quantization precision of the IRS and the ADCs should respectively satisfy
	\begin{align}\label{inequ_B}
	B \ge {\log _2}\pi  - {\log _2} {\rm arcsinc}\sqrt {\frac{{\hat a}}{{P{{\left| {\alpha \beta } \right|}^2}{N^{\rm{2}}}\left( {{2^\delta }\gamma M - \hat a + \hat a\gamma } \right)}}},
	\end{align}
	\begin{align}\label{inequ_gamma}
	{ \rho \le  1-f(\hat s)=1-}\frac{{\left( {\left( {1 + M\hat s} \right){2^{ - \delta }} - 1} \right)\left( {1 + \hat s} \right)}}{{\left( {\left( {1 + M\hat s} \right){2^{ - \delta }} - 1 + M} \right)\hat s}},
	\end{align}
	where ${{\hat a}} = 1 + \frac{{\gamma P{{\left| {\alpha \beta } \right|}^2}{N^{\rm{2}}}M}}{{1 + \left( {1 - \gamma } \right)P{{\left| {\alpha \beta } \right|}^2}{N^{\rm{2}}}}} - {2^\delta }$, $\hat s = P{\left| {\alpha \beta } \right|^2}{N^{\rm{2}}}{{\mathop{\rm sinc}\nolimits} ^{\rm{2}}}\left( {\frac{\pi }{{{2^B}}}} \right)$, { the mapping from $\rho$ to $b$ is given in Table \ref{tab1} and $\left(1-f\left(\hat s\right)\right)$ is a decreasing} function with respect to $N$, $P$, $B$ and $M$.
\end{proposition}

\itshape {Proof:}  \upshape See Appendix B. \hfill $\Box$

On one hand, Proposition \ref{proposition1} reveals that the precision of ADCs  at BSs should increase with $N$, $P$, $B$ and $M$ to guarantee the acceptable rate loss. For $N\to\infty$ or $P\to\infty$, we have $\rho\le0$ based on (\ref{inequ_gamma}). For the increase of $M$,  the decreasing speed of $\rho$ decreases based on (\ref{appBinsight}) which shows that the first-order derivative is approaching zero when $M$ is very large.
 
On the other hand, Proposition \ref{proposition1} shows that $B\ge0$  when $N\to\infty$ or $P\to\infty$, which means that the precision of IRS phase shifts is insignificant under large $N$ or $P$. Besides, when $\delta=0$, we have $B\ge\infty$ and $\rho\le0$. These  results, which are consistent with the Remark \ref{remark11} and Remark \ref{remark22},  verify the accuracy of  Proposition \ref{proposition1}.

\begin{remark}
Assume that the transmission power $P$ is scaled with $M$ and $N$ according to $P = \frac{{{E_u}}}{{M{N^2}}}$, where $E_u$ is fixed. When $M \to \infty $, we can obtain
\begin{align}\label{remark3}
R\mathop  \to \limits^{a.s.} {\log _2}\left( {1 + \gamma {E_u}{{\left| {\alpha \beta } \right|}^2}{{\rm sinc}^2}\left( \frac{\pi}{2^B}\right)} \right).
\end{align}
This indicates that with the assistance of the IRS, the transmit power can be scaled down proportionally to $1/(MN^2)$ to achieve the same rate of a single-antenna non-IRS-aided system with transmit power ${\gamma {E_u}{{\rm sinc}^2}\left( \frac{\pi}{2^B} \right)}$ and channel strength $\alpha \beta $.
\end{remark}

\vspace{-10pt}
\section{Numerical Results}
In this section, we  validate our analytical results through simulations. Unless otherwise stated, our parameters are set as $N=64$, $M=16$, $B=1$, $b=2$, { $\alpha^2=10^{-3}d_1^{-2.2}$, $\beta^2=10^{-3}d_2^{-2.2}$ \cite{9110849}, $d_1=10$ m, $d_2=40$ m, $P=\frac{20\,{\rm dbm}}{-80\,{\rm dbm}}$.} { Theoretical results are calculated using (\ref{rateExp}) while  simulation results are obtained by averaging over 100 independent calculations of (\ref{rate}).}
\begin{figure*}
	\setlength{\abovecaptionskip}{-5pt}
	\setlength{\belowcaptionskip}{-15pt}
	\centering
	\begin{minipage}[t]{0.33\linewidth}
		\centering
		\includegraphics[width=2.3in]{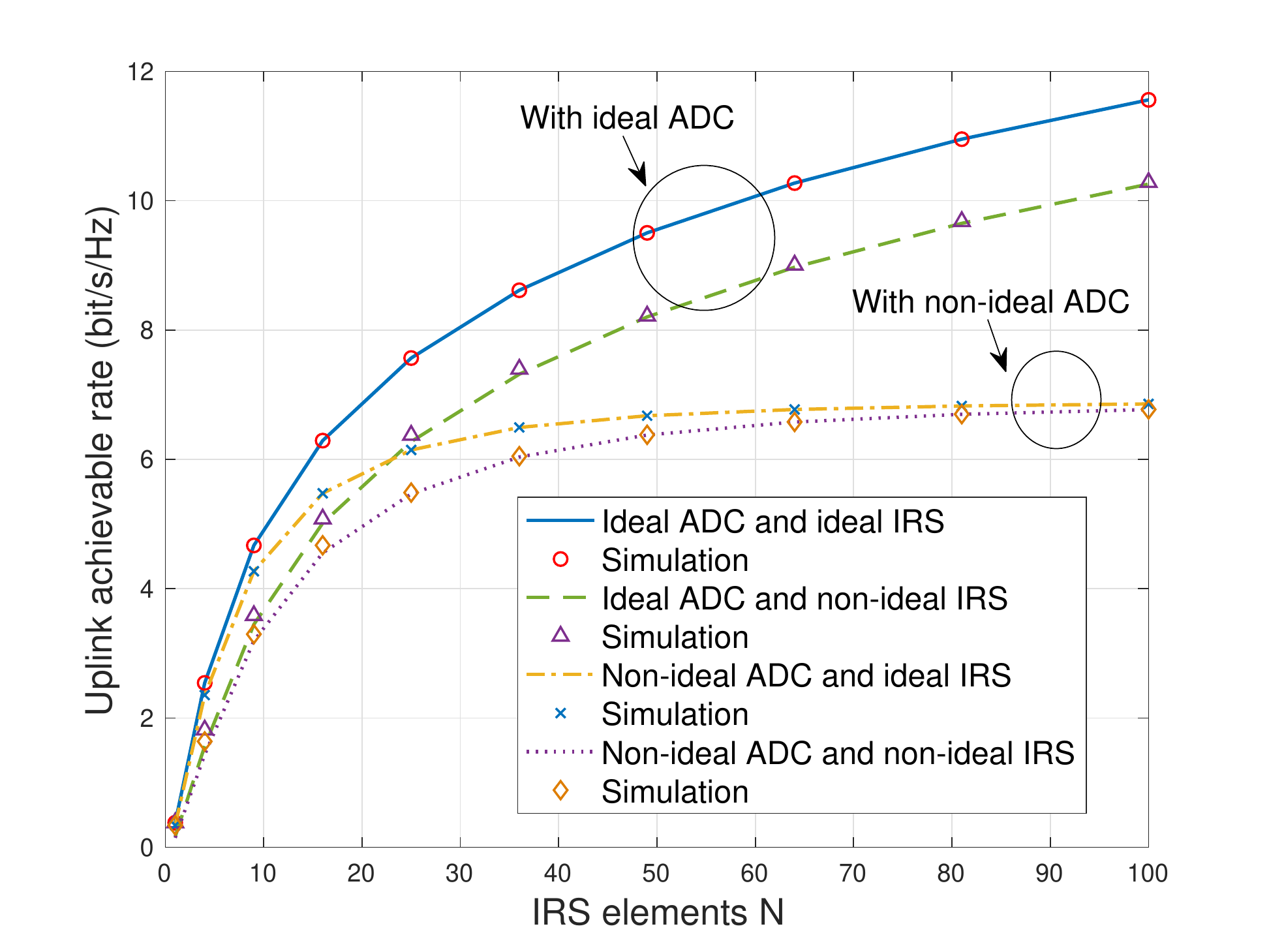}
		\caption{Uplink achievable rate versus $N$.}
		\label{figure1}
	\end{minipage}%
	\begin{minipage}[t]{0.33\linewidth}
		\centering
		\includegraphics[width=2.3in]{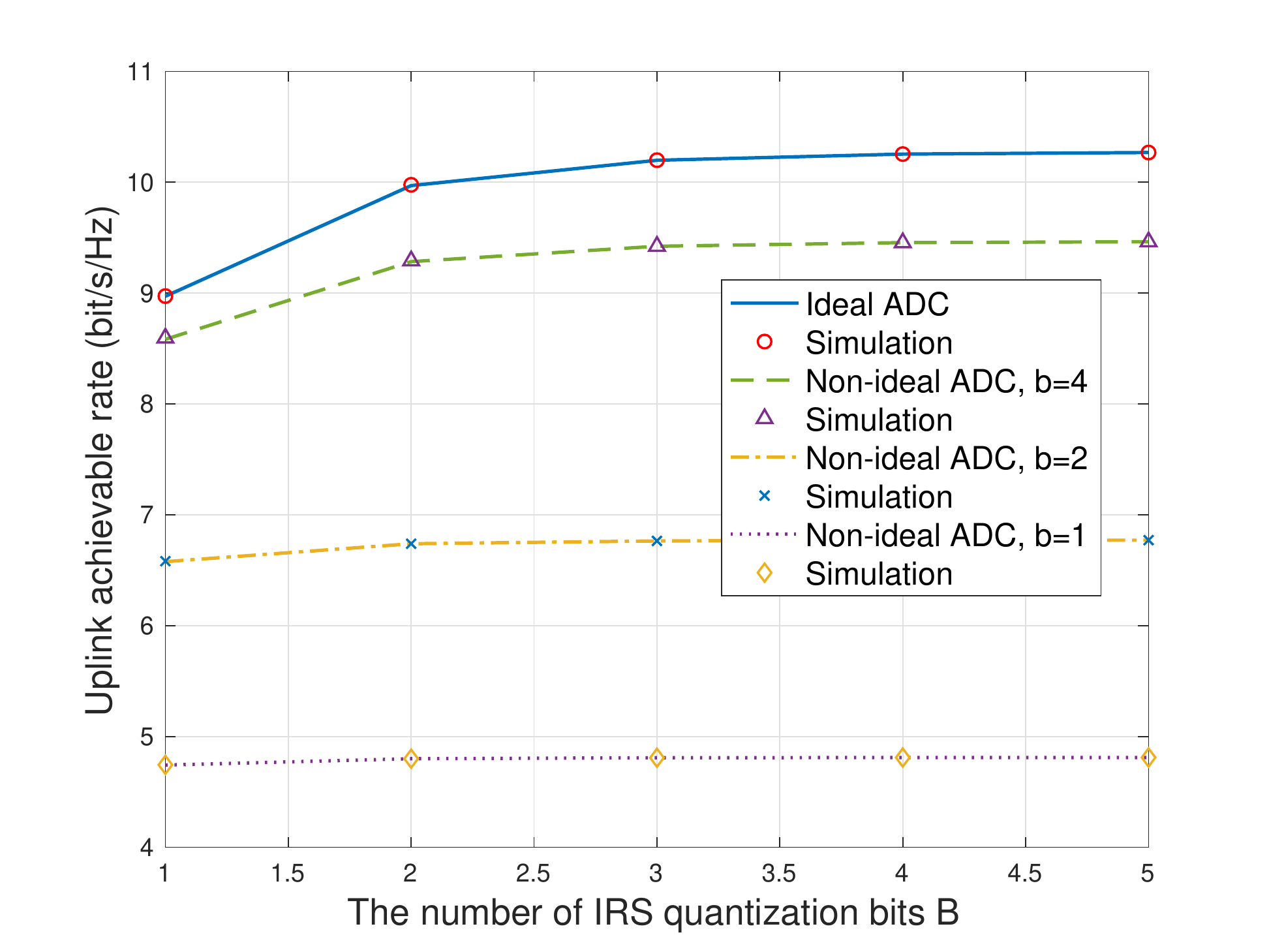}
		\caption{The impact of IRS quantization bits $B$.}
		\label{figure2}
	\end{minipage}
	\begin{minipage}[t]{0.33\linewidth}
		\centering
		\includegraphics[width=2.3in]{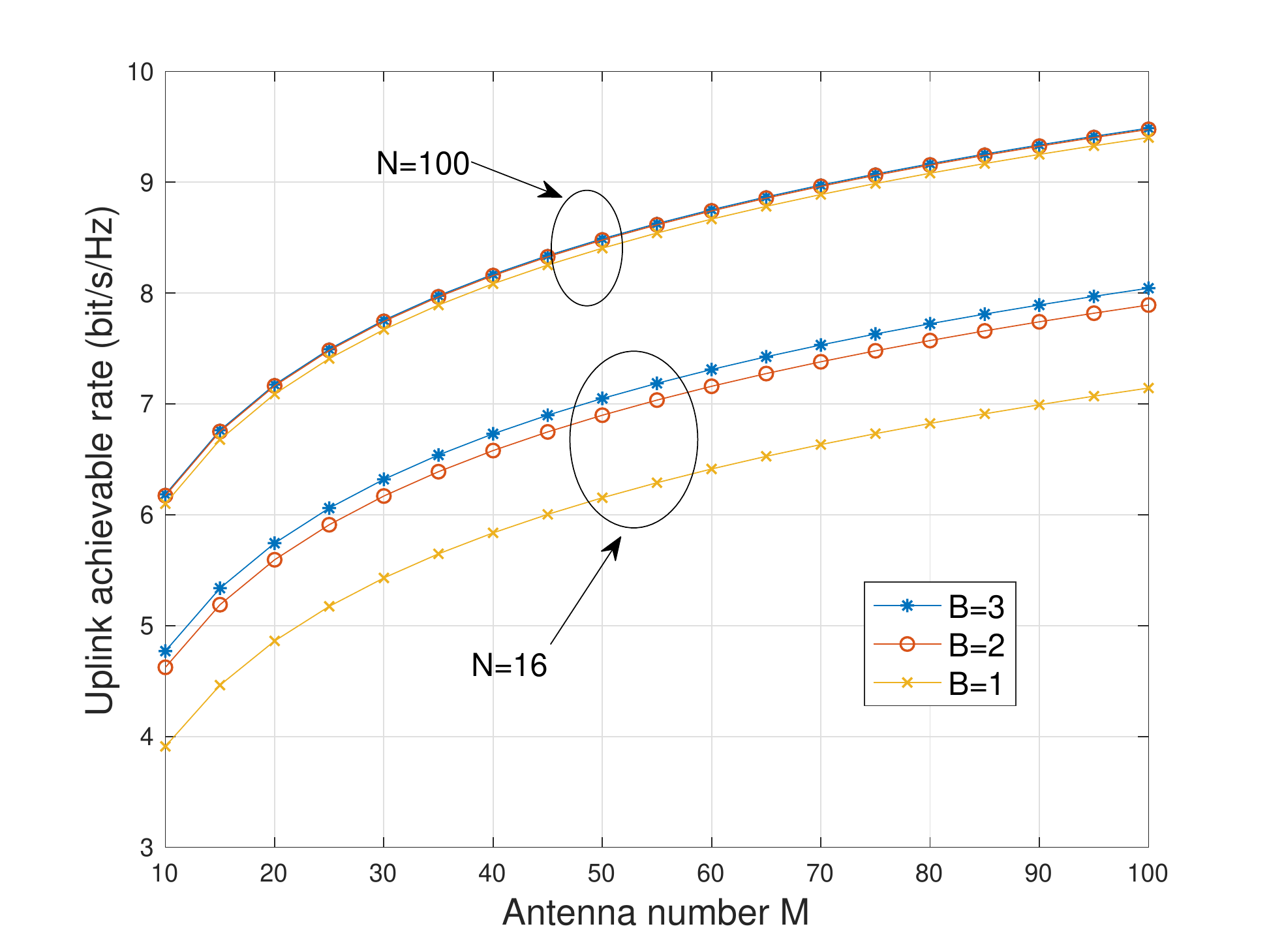}
		\caption{Rate versus the number of antennas $M$.}
		\label{figure3}
	\end{minipage}
\vspace{-10pt}
\end{figure*}


Fig. \ref{figure1} shows the the impacts of the number of  IRS reflecting elements on the uplink achievable rate under four different scenarios. It can be observed that our analytical results match well with the simulation results, which verifies the correctness of our derivations. Meanwhile, the simulation results of Fig. \ref{figure1} are consistent with our discussions in Remark 1 and Remark 2. With an ideal ADC, we can see that the rate can be increased infinitely by enlarging the size of IRS, and the gap between the ideal IRS (i.e., without phase noise) and non-ideal IRS tends to be a constant when $N$ becomes large. On the contrary, when considering the effect of the low-resolution ADC, the rate with/without phase noise  finally approaches the same upper bound.

Fig. \ref{figure2} shows the influence of the phase noise under different ADCs resolutions. We can see that phase noise at the IRS has a larger impact when the ADC has a higher resolution. When the precision of the ADC decreases, only the serious phase noise, i.e., $B=1$, will bring obvious degradation of system performance. However, even with ideal ADCs, a 3-bits quantized IRS is sufficient to achieve the near-optimal rate performance. This finding validates the feasibility of using low-quality hardware at IRS (e.g., $B=1,2$)  when the BSs are equipped with low-resolution ADCs.

Fig. \ref{figure3} shows the uplink achievable rate scaling law with respect to the number of the antennas at the BS. We can see that the achievable rate increases logarithmically with $M$ when $M$ is large. Besides, we can also see that the impact of phase noise becomes marginal when $N$ is large.

\vspace{-10pt}
\section{Conclusion}
In this paper, we have analyzed the rate performance of an uplink IRS-aided MISO system with hardware limitations at both the BS and the IRS. Using our derived expressions, we have shown that the rate scaling law is ${\mathcal O}(\log_2(M))$ which is not  influenced by the phase noise at the IRS. Our results  have also  shown that when the number of  reflecting elements of the IRS approaches infinity, the rate is only limited by the precision of the ADCs. Numerical simulations have verified the accuracy of our analytical results.

\vspace{-10pt}
\begin{appendices}
\section{}
To derive the expression of the rate in (\ref{rate}), we need to calculate ${{{\bf g}^H}\tilde {\bf g}}$, ${{{\left\| {\bf g} \right\|}^2}}$ and $  {\bf g}^H{\rm diag} \left( {P\tilde {\bf g}{{\tilde {\bf g} }^H} + {\bf I}} \right){\bf g}$, respectively. Please note that in these terms, the only random variable is phase noise ${\hat \theta }_n$, which will converge to its expectation when $N\to\infty$ due to the law of large number.

First, by using the optimal phase shift design in (\ref{optimal_ps}), the cascaded channels with and without phase noise can be respectively expressed as
\begin{align}\label{prof1}
\begin{array}{l}
{\bf g} = {\bf H_2} {\bf \Theta} {\bf h_1} = \alpha \beta N{{\bf a}_M}\left( {\phi _r^a,\phi _r^e} \right),\\
\tilde {\bf g} = {\bf H_2}\tilde {\bf \Theta} {\bf h_1} = \alpha \beta {{\bf a}_M}\left( {\phi _r^a,\phi _r^e} \right)\sum_{n = 1}^N {{e^{j{{\hat \theta }_n}}}} .
\end{array}
\end{align}

Then using the result in (\ref{prof1}), we have
{\begin{align}\label{prof2}
&{\left\| {\bf g} \right\|^2} = \mathbf{g}^{H} {\mathbf{g}}\!=\!|\alpha \beta|^{2} N^2  {\left\|\mathbf{a}_{M}\left(\phi_{r}^{a}, \phi_{r}^{e}\right)\right\|^2}\!=\! {\left| {\alpha \beta } \right|^2}{N^2}M,\\
&\mathbf{g}^{H} \tilde{\mathbf{g}}=\!|\alpha \beta|^{2} N {\left\|\mathbf{a}_{M}\left(\phi_{r}^{a}, \phi_{r}^{e}\right)\right\|^2} \!\sum\limits_{n=1}^{N} \!e^{j \hat{\theta}_{n}}\!=\!|\alpha \beta|^{2} N M \!\sum\limits_{n=1}^{N} \!e^{j \hat{\theta}_{n}}.
\end{align}}

For large $N$, by using strong law of large numbers and the continuous mapping theorem, we can obtain
\begin{align}\label{prof_sinc}
\frac{1}{N}\!\sum_{n = 1}^N \!{{e^{\pm j{{\hat \theta }_n}}}} \mathop  \to \limits^{a.s.} \mathbb{E}\!\left[ {{e^{\pm j{{\hat \theta }_n}}}} \right] \!{\mathop  = \limits^{\left( a \right)} }  \mathbb{E}\!\left[ {\cos\! \left( {{{\hat \theta }_n}} \right)} \right]\! {\mathop  = \limits^{\left( b \right)} } {\mathop{\rm sinc}\nolimits} \!\left( \frac{\pi}{2^B} \!\right),
\end{align}
where $(a)$ applies the symmetry of the odd function $\sin\left( {{{\hat \theta }_n}} \right)$ with ${\hat \theta _n} \sim {\mathcal U} \left[   \frac{-\pi}{2^B}, \frac{\pi}{2^B}\right]$, and $(b)$ is calculated based on the probability density function of ${\hat \theta }_n$. Therefore, we have
\begin{align}\label{prof3}
{{{\bf g}^H}\tilde {\bf g}} { \mathop  \to \limits^{a.s.}} {{{\left| {\alpha \beta } \right|}^2}{N^2}{M}}{{{{\mathop{\rm sinc}\nolimits} }^{\rm{}}}\left( \frac{\pi}{2^B} \right)}.
\end{align}


Then we calculate the term $  {\bf g}^H{\rm diag} \left( {P\tilde {\bf g}{{\tilde {\bf g} }^H} + {\bf I}} \right){\bf g}$. The $m$th diagonal element of ${\rm diag} \left( {P\tilde {\bf g}{{\tilde {\bf g} }^H} + {\bf I}} \right)$ can be written as
\begin{align}\label{prof5}
{\left[ {{\rm diag}\left( {P\tilde {\bf g}{{\tilde {\bf g} }^H} \!+\! {\bf I}} \right)} \right]_{mm}} \!=\! 1 \!+\! P{\left| {\alpha \beta } \right|^2}\sum\limits_{n = 1}^N {{e^{j{{\hat \theta }_n}}}} \!\!\sum\limits_{n = 1}^N {{e^{ - j{{\hat \theta }_n}}}} .
\end{align}

Thus, by applying (\ref{prof_sinc}), we have
\begin{align}\label{prof_diag_gamma}
&{\bf g}^H{\rm diag} \left( {P\tilde {\bf g}{{\tilde {\bf g} }^H} + {\bf I}} \right){\bf g}={\left| {\alpha \beta } \right|^2}{N^2}M\left[ {{\rm diag}\left( {P\tilde {\bf g}{{\tilde {\bf g} }^H} + {\bf I}} \right)} \right]_{mm}\nonumber\\
&\mathop  \to \limits^{a.s.} {\left| {\alpha \beta } \right|^2}{N^2}M\left( {1 + P{{\left| {\alpha \beta } \right|}^2}{N^{\rm{2}}}{{{\mathop{\rm sinc}\nolimits} }^{\rm{2}}}\left( \frac{\pi}{2^B} \right)} \right).
\end{align}

Finally, substituting (\ref{prof2}),  (\ref{prof3}) and (\ref{prof_diag_gamma}) into (\ref{rate}) and with some simple manipulations, we can complete the proof.

\section{}
In order to ensure that the performance loss is lower than $\delta$, for the quantization bits of IRS (i.e., $B$) and the resolution of ADCs (i.e., $\gamma$), it should respectively satisfy the following inequalities
\begin{align}\label{inequlities12}
\begin{array}{l}
{\log _2}\left( {1 + \frac{{\gamma P{{\left| {\alpha \beta } \right|}^2}{N^{\rm{2}}}M}}{{\gamma  + \left( {1 - \gamma } \right)\left( {1 + P{{\left| {\alpha \beta } \right|}^2}{N^{\rm{2}}}} \right)}}} \right) - R\left( {\gamma ,B} \right) \le \delta ,\\
{\log _2}\left( {1 + P{{\left| {\alpha \beta } \right|}^2}{N^{\rm{2}}}M{{{\mathop{\rm sinc}\nolimits} }^{\rm{2}}}\!\left( {\frac{\pi }{{{2^B}}}} \right)} \right) - R\left( {\gamma ,B} \right) \le \delta ,
\end{array}
\end{align}
where $R\left( {\gamma ,B} \right)$ is the general rate expression (\ref{rateExp}) presented in Theorem \ref{thm1}.

By solving  inequalities in (\ref{inequlities12}), we can obtain the constraints of $B$ and $\rho$ in (\ref{inequ_B}) and (\ref{inequ_gamma}), respectively.

Next, based on the expression of $f(\hat s)$, the first-order derivative of $f(\hat s)$ with respect to $\hat s$ and $M$ can be respectively written as
%
\begin{align}
&\frac{{\partial f\left( {\hat s} \right)}}{{\partial \hat s}}\mathop  = \limits^{\left( c \right)} \frac{{2M\hat s{2^{ - \delta }}\left( {1 - {2^{ - \delta }}} \right) + \left( {M - 1} \right)\left( {1 - {2^{ - \delta }}} \right)}}{{{{\left( {\hat s{2^{ - \delta }} + M{{\hat s}^2}{2^{ - \delta }} - \hat s + M\hat s} \right)}^2}}}\nonumber\\
&\qquad\quad+ \frac{{{M^2}{{\hat s}^2}{2^{ - \delta }}\left( {1 - {2^{ - \delta }}} \right) + {2^{ - \delta }}\left( {1 - {2^{ - \delta }}} \right)}}{{{{\left( {\hat s{2^{ - \delta }} + M{{\hat s}^2}{2^{ - \delta }} - \hat s + M\hat s} \right)}^2}}} \ge 0,\\
&\frac{{\partial f\left( {\hat s} \right)}}{{\partial M}}\mathop  = \limits^{\left( c \right)} \frac{{\hat s\left( {1 - {2^{ - \delta }}} \right) + {{\hat s}^2}\left( {1 - {2^{ - \delta }}} \right)}}{{{{\left( {\hat s{2^{ - \delta }} + M{{\hat s}^2}{2^{ - \delta }} - \hat s + M\hat s} \right)}^2}}} \ge 0,\label{appBinsight}
\end{align}
where $(c)$ is calculated based on the derivative law of fractions and through some algebraic simplifications.


Since $\hat s$ is increasing with $N$, $P$ and $B$, thus we can complete the proof.
\end{appendices}

\bibliographystyle{IEEEtran}
\vspace{-6pt}
\bibliography{myref.bib}

\end{document}